\documentclass[conference,pdftex]{IEEEtran}
\usepackage{graphicx}
\usepackage{xcolor}
\usepackage{amsmath}
\usepackage{amssymb}
\usepackage[nobreak]{cite}
\usepackage{seqsplit}
\usepackage{colortbl}
\usepackage{xspace}
\usepackage{url}

\newcommand{\Mod}[1]{\texttt{#1}}
\newcommand{\Bug}[1]{\textsf{#1}}
\newcommand{\BLT}[1]{\textsf{#1}}

\newcommand{\ie}{\textit{i.e.}}
\newcommand{\etal}{\textit{et al.}\xspace}

\newcommand{\Heading}[1]{\subsubsection{#1}}
\newcommand{\RQ}[1]{\textit{RQ}${}_{\mathrm{#1}}$}
\usepackage{framed}
\newcommand{\Conclusion}[1]{\begin{framed}\noindent #1\end{framed}}
\newcommand{\G}{\cellcolor{lightgray}}
\newcommand{\B}{\bf}

\begin{document}

\title{Large-Scale Evaluation of Method-Level\\Bug Localization with FinerBench4BL}

\author{%
  \IEEEauthorblockN{Shizuka Tsumita}
  \IEEEauthorblockA{%
  \textit{Tokyo Institute of Technology}\\
  Tokyo 152--8550, Japan\\
  tsumita@se.c.titech.ac.jp}
  \and
  \IEEEauthorblockN{Shinpei Hayashi}
  \IEEEauthorblockA{%
  \textit{Tokyo Institute of Technology}\\
  Tokyo 152--8550, Japan\\
  hayashi@c.titech.ac.jp}
  \and
  \IEEEauthorblockN{Sousuke Amasaki}
  \IEEEauthorblockA{%
  \textit{Okayama Prefectural University}\\
  Okayama 700--0961, Japan\\
  amasaki@cse.oka-pu.ac.jp}
}

\maketitle
\thispagestyle{plain}

\begin{abstract}
  Bug localization is an important aspect of software maintenance because it can locate modules that need to be changed to fix a specific bug.
  Although method-level bug localization is helpful for developers, there are only a few tools and techniques for this task; moreover, there is no large-scale framework for their evaluation.
  In this paper, we present FinerBench4BL, an evaluation framework for method-level information retrieval-based bug localization techniques, and a comparative study using this framework.
  This framework was semi-automatically constructed from Bench4BL, a file-level bug localization evaluation framework, using a repository transformation approach.
  We converted the original file-level version repositories provided by Bench4BL into method-level repositories by repository transformation.
  Method-level data components such as oracle methods can also be automatically derived by applying the oracle generation approach via bug-commit linking in Bench4BL to the generated method repositories.
  Furthermore, we tailored existing file-level bug localization technique implementations at the method level.
  We created a framework for method-level evaluation by merging the generated dataset and implementations.
  The comparison results show that the method-level techniques decreased accuracy whereas improved debugging efficiency compared to file-level techniques.
\end{abstract}
\begin{IEEEkeywords}
  bug localization, information retrieval, repository transformation
\end{IEEEkeywords}

\section{Introduction}\label{s:introduction}

Bug localization is the process of identifying the location of a bug.
Developers must fix many bugs in large-scale software projects, and debugging software is difficult and time-consuming~\cite{whyProgramsFail}.
As this can be a tedious task in large-scale software development projects, numerous ideas have been proposed to automate this process using software development information.
For instance, we can identify the locations of a bug using the description of bug reports, \ie, information retrieval~(IR)-based techniques~\cite{lukins-ist2010,nguyen-ase2011}, or execution traces, \ie, dynamic analysis~\cite{Wong2014BRTracer}.
Several hybrid techniques that combine a base technique with additional information have been proposed to improve bug localization accuracy.
For example, \BLT{BugLocator}~\cite{Zhou2012BugLocator} improved an IR-based technique with similar bug reports that were previously resolved.
\BLT{BLUiR}~\cite{Saha2013BLUiR} incorporated structural information in addition to using similar bug reports.
\BLT{AmaLgam}~\cite{Wang2014AmaLgam} combined the version history, structural information, and similar bug reports.

Most existing IR-based bug localization~(IRBL) techniques follow file-level recommendations.
They output a ranked list of suspicious files; therefore, their evaluations were performed at the file level.
This granularity may be too coarse for developers to debug.
In particular, IRBL approaches may recommend large files with more than 500 lines of code.
Debugging efforts can be decreased if it is possible to recommend code fragments to be fixed at the method level.

However, in the literature, there are only a few IRBL techniques and their implementations at the method level~\cite{Razzaq2021Nsift,Zhang2019Fine,Youm2017BLIA}.
To the best of our knowledge, no study has performed a comprehensive comparison and evaluation of method-level IRBL techniques, and only IR techniques have been compared~\cite{Chakkrit2018IR}.
In addition, no publicly available evaluation framework enables comparison at the method level, and we have little knowledge of method-level IRBL approaches.

Therefore, in this study, we propose FinerBench4BL, an evaluation framework for method-level bug localization techniques.
This framework is based on Bench4BL~\cite{Lee2018Bench4BL}, an existing evaluation framework for file-level bug localization techniques.
We built a method-level bug localization dataset by applying a repository transformation to the repositories in the Bench4BL dataset and prepared method-level IRBL implementations by modifying the file-level ones.

This study also presents a performance study of method-level IRBL techniques.
The results demonstrated that method-level IRBL techniques do not improve the accuracy but reduce debugging effort compared with file-level bug localization.
We demonstrated a similar performance across different levels and revealed the need for improvements in method-level bug localization.

The main contributions of this study with respect to method-level bug localization techniques are as follows:
\begin{itemize}
  \item a transformational approach to obtain method-level IRBL techniques and datasets,
  \item construction of an evaluation framework that enables comparison of IRBL techniques at the method level, and
  \item confirmation of the superiority and inferiority between techniques at the method level, which replicates existing IRBL approaches applied to the different levels.
\end{itemize}

The remainder of this paper is organized as follows.
In the next section, we briefly introduce IR-based bug localization and its evaluation framework.
Section~\ref{s:motivation} discusses the issues in terms of module granularity, and related work is introduced in Section~\ref{s:relatedwork}.
The approach proposed in this study that leads to the solution of these issues is presented in Section~\ref{s:approach}.
Section~\ref{s:evaluation} presents the experimental setup used in the analysis and comparison of the results of the file and method levels.
Threats to validity are presented in Section~\ref{s:validity}.
Finally, the conclusions and future work are presented in Section~\ref{s:conclusion}.

%%%%%%%%%%%%%%%%%%%%%%%%%%%%%

\section{Preliminaries}\label{s:preparation}

\subsection{IR-Based Bug Localization}

Bug localization is the process of finding buggy locations in source code based on the information about a given bug.
This study explicitly focuses on IR-based bug localization~(IRBL), which uses textual analysis to determine bug locations.
IRBL techniques use a bug report and source code as inputs and output a ranked list of modules to be fixed.
Some IRBL techniques may use past bug reports, stack traces, or historical information as additional inputs.
The given bug report serves as a query for searching the given source code.
A bug report includes the bug ID, summary, dates when the bug is opened or closed, and detailed description.
The description of a bug may include stack traces that are extracted and utilized using advanced IRBL techniques.
Finally, IRBL techniques compute the similarity score between the bug report and source code files considering the additional information and output a ranked list of source files on the score.

\begin{table}[tb]\centering
  \caption{Applying \BLT{BugLocator} to \Bug{CODEC-199}}\label{t:ranklist_eg}
  \begin{tabular}{rlr} \hline
    Rank & Filename & Score \\\hline
    1 &\G\Mod{Soundex.java}               & 0.800 \\
    2 &  \Mod{DaitchMokotoffSoundex.java} & 0.618 \\
    3 &  \Mod{RefinedSoundex.java}        & 0.564 \\
    4 &  \Mod{SoundexTest.java}           & 0.500 \\
    5 &  \Mod{RefinedSoundexTest.java}    & 0.388 \\\hline
  \end{tabular}
\end{table}

As an IRBL example, an excerpt result of applying \BLT{BugLocator}~\cite{Zhou2012BugLocator} to bug \Bug{CODEC-199}\footnote{https://issues.apache.org/jira/browse/CODEC-199} is shown in Table~\ref{t:ranklist_eg}.
The columns indicate the rank, filename, and similarity score of the target file for the query bug report.
The highlighted row, \Mod{Soundex.java}, specifies the \emph{oracle}, \ie, the buggy file for this bug.
The higher the rank of the file, the more likely it is to contain a bug.
Therefore, developers search for bugs starting with \Mod{Soundex.java} in the table.

\subsection{IRBL Evaluation Framework}

Researchers often employ a retrospective approach by applying IRBL techniques to previously resolved bugs to evaluate their performance.
When resolving a bug, the list of fixed modules is regarded as the oracle list of modules that should be localized associated with the bug.
A set of resolved bug reports and their associated lists of fixed modules form an IRBL evaluation dataset.
The performance of IRBL techniques can be evaluated by comparing the oracle list of modules to the ranked list of modules produced by the techniques.

Bench4BL\footnote{https://github.com/exatoa/Bench4BL}\cite{Lee2018Bench4BL} is a large-scale framework proposed by Lee \etal\ to evaluate the performance of file-level bug localization techniques.
Resolved bug reports obtained from the issue tracking systems for 46 projects and their lists of oracle source files to be localized were collected and provided as a dataset for evaluation.
In addition to the dataset, they have attached the implementations of \BLT{BugLocator}~\cite{Zhou2012BugLocator}, \BLT{BLUiR}~\cite{Saha2013BLUiR}, \BLT{BRTracer}~\cite{Wong2014BRTracer}, \BLT{AmaLgam}~\cite{Wang2014AmaLgam}, \BLT{BLIA}~\cite{Youm2015BLIA}, and \BLT{Locus}~\cite{Wen2016Locus}.

In Bench4BL, released versions of projects are regarded as source code snapshots to be searched.
Each version was created by checking out files from a Git repository in the dataset.
An oracle list of files corresponding to a bug report was generated based on the commit history in a repository.
Once a bug-fixing commit is detected based on the matching between its commit message and the ID of the bug, the changed files in the commit are regarded as files to be fixed to resolve the bug.
In addition, \BLT{AmaLgam} and \BLT{BLIA} utilize the historical information (also obtained from the Git repository) to improve the bug localization accuracy.
In summary, all information used in applying IRBL techniques follows the original Git repositories.

%%%%%%%%%%%%%%%%%%%%%%%%%%%%%

\section{Motivation}\label{s:motivation}

Most existing IR-based bug localization techniques localize buggy code at the file level.
Techniques such as \BLT{BugLocator}~\cite{Zhou2012BugLocator}, \BLT{BLUiR}~\cite{Saha2013BLUiR}, \BLT{AmaLgam}~\cite{Wang2014AmaLgam}, and \BLT{BLIA}~\cite{Youm2015BLIA} all recommend suspicious modules at the file level.
This section presents the challenges of file-level bug localization and method-level bug localization.

\subsection{Challenges of File-Level Bug Localization}

When recommending buggy files, IR-based bug localization techniques may produce very large files that contain methods unrelated to the bug, making it challenging to identify bug locations.
For example, consider the bug \Bug{CODEC-221}\footnote{https://issues.apache.org/jira/browse/CODEC-221}.
For this bug, the three methods in \Mod{HmacUtils.java} must be fixed.
This file is 729 lines long and includes 40 methods.
Therefore, significant time is required to identify the bug location in the entire file.
It would be more helpful if the three buggy methods were recommended to be fixed directly.

\begin{table*}[tb]\centering
  \caption{Applying \BLT{BLIA} at File and Method Levels to \Bug{CODEC-221}}\label{t:rankloc}
  \begin{tabular}{r|lrr|lrr} \hline
    \multicolumn{1}{c|}{} & \multicolumn{3}{c|}{File level} & \multicolumn{3}{c}{Method level} \\
     Rank & Module                         & Score & LOC & Module                                            & Score & LOC \\\hline
     1 &\G\Mod{HmacUtils.java}             & 0.640 & 729 &  \Mod{BaseNCodecInputStream\#reset()}         & 0.640 & 15 \\
     2 &  \Mod{DigestUtils.java}           & 0.251 & 752 &\G\Mod{HmacUtils\#updateHmac(Mac,InputStream)} & 0.573 & 29 \\
     3 &  \Mod{HmacUtilsTest.java}         & 0.095 & 237 &\G\Mod{HmacUtils\#updateHmac(Mac,byte[])}      & 0.565 & 19 \\
     4 &  \Mod{BaseNCodecInputStream.java} & 0.048 & 184 &\G\Mod{HmacUtils\#updateHmac(Mac,String)}      & 0.565 & 19 \\\hline
  \end{tabular}
\end{table*}

\begin{table*}[tb]\centering
  \caption{Comparison of IRBL Studies}\label{t:relatedwork}
  \begin{tabular}{l|ccccc} \hline
       & Additional information & Method level & \# techniques & \# projects & Multi-version \\\hline
    Lee \etal~(Bench4BL)\cite{Lee2018Bench4BL}       & $\checkmark$ &  & 6 & 46 & $\checkmark$ \\
    Youm \etal~(\BLT{BLIA1.5})\cite{Youm2017BLIA}  & $\checkmark$ & $\checkmark$ & 1 & 3 &  \\
    Amasaki \etal~\cite{Amasaki2020}                 & $\checkmark$ & $\checkmark$ & 1 & 43 & $\checkmark$ \\
    Razzaq \etal~(\BLT{BoostNSift})\cite{Razzaq2021Nsift} & $\checkmark$ & $\checkmark$ & 4 & 4 &  \\
    Chakkrit \etal~\cite{Chakkrit2018IR}               &  & $\checkmark$ & 4 & 2 &  \\
    Our Approach                                       & $\checkmark$ & $\checkmark$ & 5 & 37 & $\checkmark$ \\\hline
  \end{tabular}
\end{table*}

Table~\ref{t:rankloc} lists the results of applying \BLT{BLIA} to this bug report at both file and method levels.
This table shows the names of the modules recommended by \BLT{BLIA} with their ranks at both the file and method levels.
The score represents the points used to rank each file, and LOC shows the lines of code of the module.
Rows with the correct answers are highlighted.
In this example, the correct answer at file level, \Mod{HmacUtils.java}, is recommended at the first rank with a score of 0.640.
However, identifying the methods that need to be fixed in this file, which consists of more than 700 lines, is challenging.
Conversely, all buggy methods are localized up to the fourth in the list at the method level, and the sum of their LOC is only 82.
The use of method-level IRBL reduces the effort required to read the 729 lines of \Mod{HmacUtils.java} by 11\%.

In addition, the ratio of buggy methods to the total number of methods in a buggy file is generally small.
In the context of bug prediction, Hata \etal\ investigated the ratio of buggy methods in a target project to ascertain the effectiveness of fine-grained bug predictions \cite{Hata2012}.
The results showed that the median number of buggy methods was 1--2, whereas that of all methods was 8--22.
Therefore, file-level recommendations may be inefficient in identifying bug locations.

\subsection{Challenges of Method-Level Bug Localization}

Currently, there is a lack of knowledge regarding method-level IRBL because only a few method-level techniques exist.
To the best of our knowledge, only a few bug localization techniques, such as \BLT{BLIA1.5}\cite{Youm2017BLIA}, \BLT{FineLocator}\cite{Zhang2019Fine}, or \BLT{BoostNSift}\cite{Razzaq2021Nsift}, work at the method level.
Furthermore, there is no unified framework for evaluating techniques at the method level, and the performance differences between the granularity levels and techniques remain unclear.
Therefore, it is necessary to establish an evaluation framework for various techniques.

%%%%%%%%%%%%%%%%%%%%%%%%%%%%%
\section{Related Work}\label{s:relatedwork}

\subsection{IRBL Approaches}

To date, many techniques have been studied that recommend bug locations based on information retrieval.
Some of these techniques add other information to the similarity between the source code and bug reports.
\BLT{BugLocator}\cite{Zhou2012BugLocator} uses similarity to previous bug reports and file size.
\BLT{BLUiR}\cite{Saha2013BLUiR} uses structural information of source code.
\BLT{BRTracer}\cite{Wong2014BRTracer} uses the stack trace information in bug reports.
\BLT{AmaLgam}\cite{Wang2014AmaLgam} and \BLT{Locus}\cite{Wen2016Locus} use historical data.
\BLT{BLIA}\cite{Youm2015BLIA} combines all of these to localize bugs.
The aforementioned techniques localize bugs at the file level and were evaluated with limited data, such as old versions of JDT or AspectJ\@.
After that, Bench4BL, an evaluation framework proposed by Lee \etal\cite{Lee2018Bench4BL}, evaluated these six techniques for 46 projects at the file level.
They suggested that IRBL techniques should be evaluated more accurately using multiple-version matching, which searches for a version in which a bug report is submitted.

Method-level IRBL is considered to aid developers in debugging more, and several method-level IRBL techniques have been proposed.
\BLT{BLIA1.5} proposed by Youm \etal\ extends \BLT{BLIA} for method level bug localization\cite{Youm2017BLIA}.
Amasaki \etal\cite{Amasaki2020} tailored \BLT{BLUiR} to localize bugs at the method level by using information outside the method.
\BLT{BoostNSift} was proposed by Razzaq \etal\cite{Razzaq2021Nsift} to filter source code based on bug report text.

Each method-level bug localization technique was evaluated using different datasets.
Youm \etal\ used AspectJ, SWT, and ZXing, whereas Amasaki \etal\ used a part of the Bench4BL dataset.
Razzaq \etal\ compared \BLT{BoostNSift} with \BLT{BLUiR}, \BLT{BugLocator}, and \BLT{BLIA}, and used the dataset used by Youm \etal\ plus Eclipse.
Numerous method-level techniques were not applied to large projects in these evaluations, and the datasets were not uniform.

However, only IR techniques without additional information have been evaluated under a unified framework at the method level.
Chakkrit \etal\ investigated the effectiveness of four IR techniques, VSM, LDA, LSI, and Entity Metric, at the method level\cite{Chakkrit2018IR}.
They proposed a metric named top-$k$ LOC, the percentage of bug reports for which at least one buggy file is found in the top-ranked files with a cumulative sum of $k$ lines of code, to investigate the performance of the techniques.
They found that the settings, such as how texts are pre-processed and which part of the texts in bug reports are used, significantly impact the performance of method-level IR techniques.
In addition, settings with good performance produce good results at any granularity.
They also stated that performance evaluation using LOC is necessary because debugging developers' efforts to find bugs may differ, even if they show similar accuracy.

Table \ref{t:relatedwork} shows the relationship between this study and previous studies.
In this table, the column of the additional information indicates whether the techniques consider other information.
The column of the method level shows whether the approach is evaluated at the method level.
The column of the multi-version shows whether the techniques localize bugs with multiple-version matching.

\begin{table}[tb]\centering
  \caption{Comparison of IRBL Datasets}\label{t:datasets}
  \begin{tabular}{llrr} \hline
    Name & Granularity & \# projects & \# bugs \\\hline
    iBUGS \cite{dallmeier2007extraction} & File   &  3 &    390 \\
    MoreBugs \cite{rao2013morebugs}      & File   &  2 &    902 \\
    BugLinks \cite{sisman2013assisting}  & File   &  2 &  5,046 \\
    Bench4BL \cite{Lee2018Bench4BL}      & File   & 46 &  9,459 \\
    Bugzbook \cite{akbar2020large}       & File   & 29 & 21,253 \\
    FDS\cite{oscar-bugdesc}              & File   & 11 &  4,429 \\
    MDS\cite{oscar-bugdesc}              & Method & 14 &    360 \\
    FinerBench4BL                        & Method & 37 &  3,344 \\\hline
  \end{tabular}
\end{table}

\subsection{IRBL Datasets}

Bug localization techniques are evaluated by comparing the list of modules produced by applying the techniques to bug reports resolved previously with the oracle list of modules that have actually been fixed when bugs were resolved.
Several sets of resolved bug reports and oracle lists of fixed modules were packed and proposed as a bug localization evaluation dataset.
Table~\ref{t:datasets} summarizes bug localization datasets.

iBUGS~\cite{dallmeier2007extraction} is a dataset developed by Dallmeier \etal
MoreBugs~\cite{rao2013morebugs} proposed by Rao \etal\, is an extended dataset for projects adopted by iBUGS by additionally attaching version history information.
BugLinks~\cite{sisman2013assisting} is a dataset proposed by Sisman \etal\ that contains non-Java projects.
iBUGS, MoreBugs, and BugLinks are relatively small datasets, comprising only 2--3 projects.

Bench4BL is a large-scale bug localization evaluation framework proposed by Lee \etal
This dataset consisted of 46 Java projects and 9,459 bug reports.
In addition, the Bench4BL framework bundles several IRBL technique implementations so that users can easily execute the techniques for projects in the provided dataset.
Akbar \etal\ proposed BugzBook~\cite{akbar2020large}, which is another large-scale bug localization dataset that contains over 20,000 bug reports from 29 projects developed in Java, C/C++, and Python.

To the best of our knowledge, most publicly available bug localization datasets are at the file level.
As mentioned above, several existing studies on method-level bug localization internally analyzed method-level datasets.
However, these datasets are not publicly available.
We believe that the inexistence of method-level datasets and evaluation frameworks hinders the promotion of method-level bug localization research.
Note that Chappaoro \etal~\cite{oscar-bugdesc} prepared IRBL datasets at the class, file~(FDS), and method level~(MDS) as part of their study on query reformulation for bug localization.
These datasets are available upon request.

%%%%%%%%%%%%%%%%%%%%%%%%%%%%%

\section{Approach}\label{s:approach}

To address the aforementioned challenges, we used the \emph{method repositories} created by the repository transformation to construct a method-level IRBL dataset, and we modified the existing file-level IRBL implementations for the method-level ones.
We constructed an evaluation framework for method-level IRBL techniques by combining them.

IRBL techniques output a list of source code files in the order of similarity to bug reports calculated using additional information, such as the text of previous bug reports and historical data.
They were evaluated by comparing the ranked list of modules obtained by applying IRBL techniques to resolve bug reports with the oracle list of modules.
Accordingly, a method-level evaluation framework needs to fulfill the following three requirements:
\begin{itemize}
  \item it should be able to output a ranked list of methods,
  \item it should be able to link bug reports to the fixed methods, and
  \item it should be able to obtain historical information about each method.
\end{itemize}

Most IRBL techniques assume that the input bug report and source code are in the form of files and calculate their similarity.
They may consider the size or history of the source files as additional information to be used to calculate the score.
A straightforward approach to make such implementations work at the method level is to add a specific process to extract information exclusively for the method of interest against existing IRBL implementations, which require substantial engineering work.
Conversely, our idea is to prepare \emph{method files} whose contents are only of the individual methods of interest, trick existing IRBL implementations to recognize that these method files are normal source files, and have them perform on these method files.
This transforms the complexity of method-level bug localization into the cost of preparing the method files and minimizes the cost of re-implementing each IRBL implementation at the method level.
Note that such method files are naturally considered incorrect Java source code for the project as a whole, but even such fake files can be analyzed without issues in most cases because IR-based approaches do not perform deep program analysis but only superficial text analysis.
By preparing the method files using a \emph{repository transformation} mechanism and converting every source file into method files at the stage of the original Git change history, most of the automated processing provided in the existing file-level IRBL evaluation framework, such as the generation of code snapshots from a repository,  extraction of the change history of each file, and identification of oracle files based on the correspondence between bug reports and commits, could be completely reused.
This could lead to an ecosystem of IRBL evaluation frameworks with multiple levels of granularity.

\begin{figure*}[tb]\centering
  \includegraphics[width=16cm]{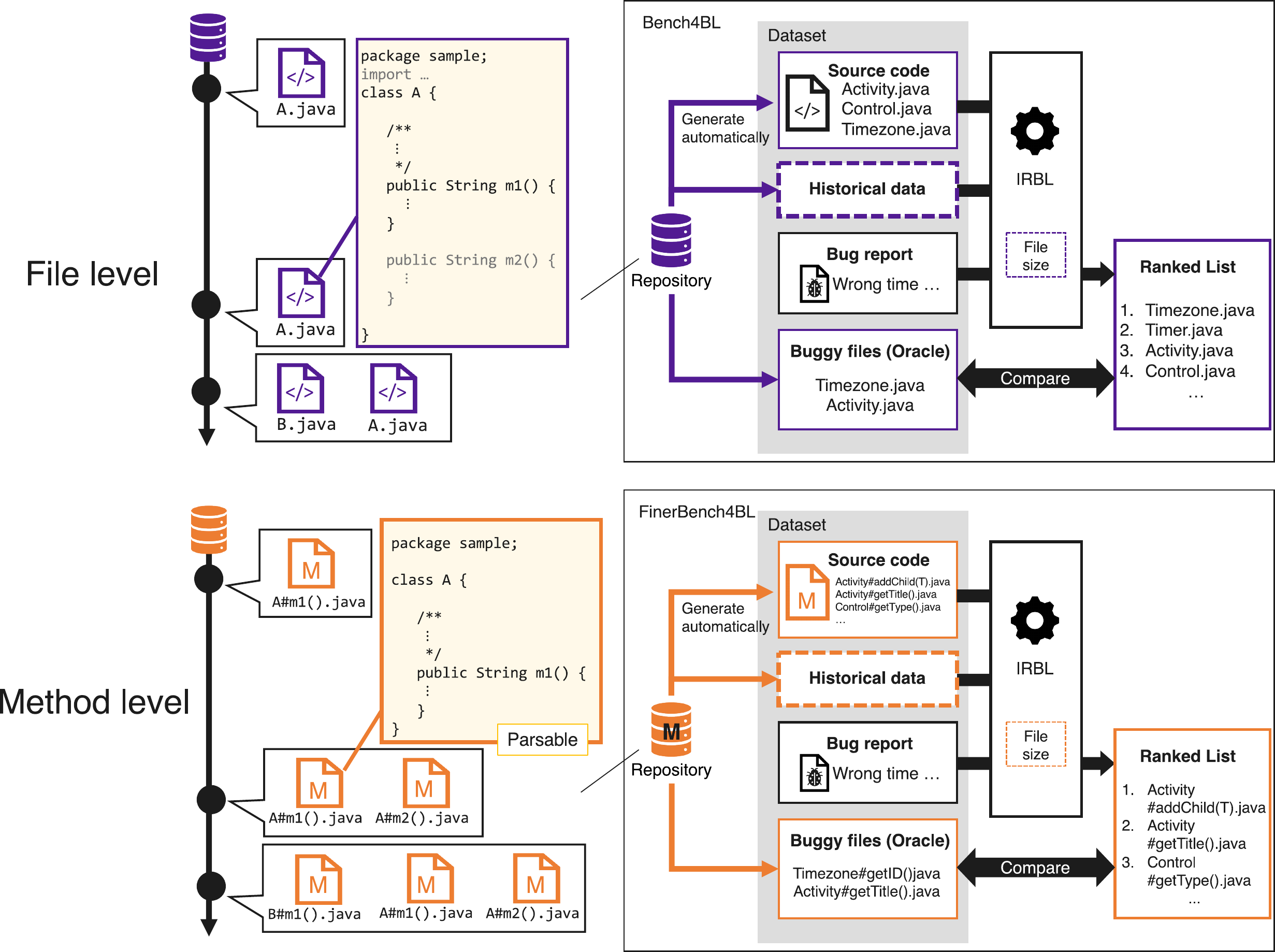}
  \caption{Overview of FinerBench4BL.}\label{f:approach}
\end{figure*}

Accordingly, we propose an approach that transforms the repository itself, which records the change history at the method level, by applying a repository transformation mechanism.
Figure~\ref{f:approach} shows the repository before and after the transformation.
As shown in the figure, the dataset repository provided by the existing evaluation framework was converted to generate source code files split by the method.
Consequently, the entire dataset was converted to a method-level dataset.
We then constructed a dataset that can compare IRBL techniques at the method level employing this method-level repository as the input. 
Fewer changes are required to modify IRBL techniques into method-level techniques.
This approach is described in more detail as follows:

\subsection{Creating Method-Level Repositories}

\begin{figure}[tb]\centering
  \includegraphics[width=\linewidth]{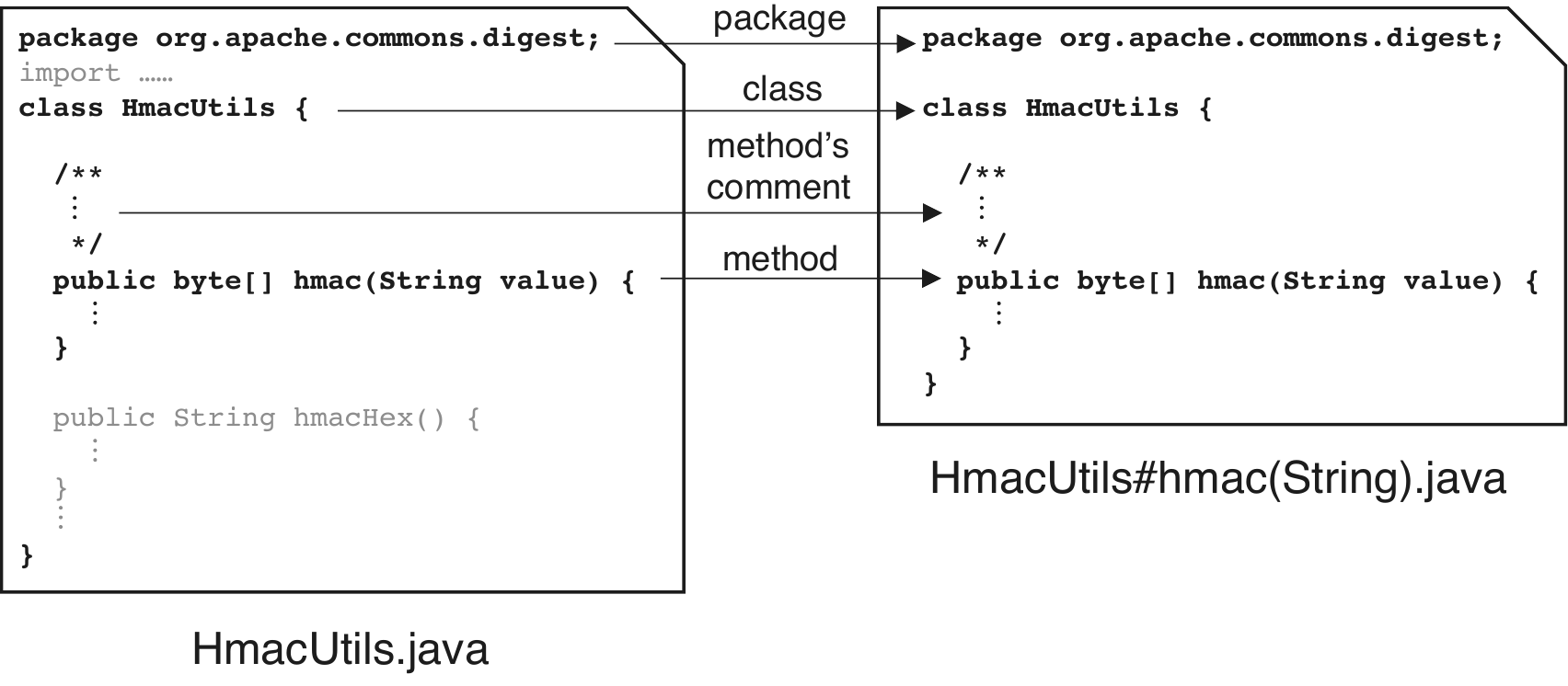}
  \caption{Example of repository transformation.}\label{f:mjava_eg}
\end{figure}

We used the Git repositories provided by Bench4BL and converted them into method-level repositories using Historinc\cite{Historinc2022}, a repository transformation tool.
We converted the target repository to the method level and obtained method files.
Figure \ref{f:mjava_eg} demonstrates an example of the splitting part of file \Mod{HmacUtils.java} in the Apache Commons CODEC\@.
By splitting the original source file, the \textit{method files}, \Mod{HmacUtils\#hmac(String).java} and \Mod{HmacUtils\#hmacHex().java}, were generated.
The package name, class name, Javadoc comment, and method body were extracted as the bodies of the method file.
Therefore, in this approach, entities outside the method, such as fields and imports of the class, were excluded from the method file.
Bugs caused outside of methods were excluded from the evaluation at the method level.

\subsection{Generating Oracles}

Bench4BL framework includes a script that links the resolved bug reports to fixed files.
This script identifies the fixed file by linking bugs to commits that resolve them by referring to the bug ID in the commit message.
Therefore, the link of bug reports can be updated by connecting bug reports to the method-level repository using the same script.
This updated link can be used for the method file to compare method-level IRBL techniques in the Bench4BL framework.

\subsection{Modifying IRBL Techniques to Method Level}

We modified the existing IRBL techniques to output a ranked list at the method level.
As explained, our approach splits source files into method files as parsable Java source code files so that the results of parsing up to the method internals can be artificially reproduced.
Therefore, if it merely uses the information available from a method file, such as file size or source code structure, no modifications are required.
Furthermore, we could obtain historical information without modifying implementations because the repository itself is already fine-grained and consists of method-level contents.
We need to modify an IRBL implementation only when it considers outside source code, such as the bug report description's stack trace, or when it requires valid source files.

For example, consider modifying \BLT{BugLocator} for method-level bug localization.
\BLT{BugLocator} outputs the ranked list of files by considering the similarity between source code and a bug report, its file size, and the similarity to past bug reports.
In this case, the original implementation could be completely reused without any changes because all the required information was available from the method file.

%%%%%%%%%%%%%%%%%%%%%%%%%%%%%

\section{Evaluation}\label{s:evaluation}

In this study, we compare and validate the method-level techniques modified by our approach and the method-level evaluation framework and answer the following research questions (RQs):
\def\RQone{How much modification is required to convert IRBL techniques to the method level?}
\def\RQtwo{How well do the method-level IRBL techniques perform?}
\begin{itemize}
  \item \RQ{1}: \RQone
  \item \RQ{2}: \RQtwo
\end{itemize}

Five of the six techniques provided by Bench4BL were used for evaluation: \BLT{BugLocator}\cite{Zhou2012BugLocator}, \BLT{BLUiR}\cite{Saha2013BLUiR}, \BLT{BRTracer}\cite{Wong2014BRTracer}, \BLT{AmaLgam}\cite{Wang2014AmaLgam}, and \BLT{BLIA}\cite{Youm2015BLIA}, excluding \BLT{Locus}\cite{Wen2016Locus}.
We excluded \BLT{Locus} because it crashed during execution, and we could not obtain the results.
We found that several techniques provided by Bench4BL had issues that prevented them from correctly calculating the similarities.
Therefore, we utilized \BLT{BLIA} implementation to imitate \BLT{BLUiR}, \BLT{AmaLgam}, and \BLT{BRTracer} by changing specific parameters to drop additional information to avoid the issues in their implementations.

\subsection{\RQ{1}: \RQone}

\Heading{Motivation}
We investigated the number of modifications to clarify whether our approach can easily convert the existing IRBL technique implementations to the method level.

\Heading{Study Design}
This study is based on the LOC of the modifications required to convert the techniques and the percentage of the total LOC of Java source files for each technique implementation.

\Heading{Results}
Figure~\ref{t:cost} illustrates the results for \RQ{1}.
We needed to modify only \BLT{BRTracer} and \BLT{BLIA} using the proposed approach.
These techniques for calculating similarity add scores to files included in the stack traces in the bug report.
We modified the process of obtaining the file name from the stack trace to obtain the method names belonging to the file.
For example, the stack trace of bug \Bug{COMPRESS-203}\footnote{https://issues.apache.org/jira/browse/COMPRESS-203} includes \Mod{\seqsplit{org.apache.commons.compress.archivers.tar.TarArchiveOutputStream.writePaxHeaders(TarArchiveOutputStream.java:485)}}.
In this case, the modified method-level techniques were needed to obtain the method name \Mod{writePaxHeaders} and add the similarity score of the corresponding method file.
It was easy to find the code location of the feature extracting the source file name from a stack trace and to modify it to extract the method file names by string manipulation, leading to only four line modifications.

\begin{table}[tb]\centering
  \caption{Changes Required to Fix Techniques for Method Level}\label{t:cost}
  \begin{tabular}{lrrr} \hline
    IRBL technique & Whole LOC & Modified LOC & Ratio (\%) \\\hline
    \BLT{BugLocator} & 3,180  & 0 & 0 \\
    \BLT{BLUiR}      & 10,469 & 0 & 0 \\
    \BLT{BRTracer}   & 10,530 & 4 & 0.038 \\
    \BLT{AmaLgam}    & 10,469 & 0 & 0 \\
    \BLT{BLIA}       & 10,474 & 4 & 0.038 \\\hline
  \end{tabular}
\end{table}

\begin{table}[tb]\centering
  \caption{Target Projects}\label{t:subject}
  {\tabcolsep=3.5pt\begin{tabular}{ll|rrrrrr}\hline
    Group  & Project & \# files & \# methods & \# versions & \# bugs \\\hline
    Commons & CODEC & 115 & 1,310 & 6 & 27 \\
 & COLLECTIONS & 525 & 6,997 & 5 & 59 \\
 & COMPRESS & 265 & 2,591 & 15 & 105 \\
 & CONFIGURATION & 447 & 6,073 & 11 & 107 \\
 & CRYPTO & 82 & 488 & 1 & 4 \\
 & CSV & 29 & 452 & 3 & 6 \\
 & IO & 227 & 2,608 & 12 & 70 \\
 & LANG & 305 & 6,336 & 15 & 158 \\
 & MATH & 1,617 & 15,695 & 15 & 175 \\
 & WEAVER & 113 & 473 & 1 & 1 \\ \hline
Jboss & ENTESB & 252 & 3,210 & 1 & 4 \\
 & JBMETA & 858 & 4,834 & 3 & 15 \\ \hline
Spring & AMQP & 408 & 3,996 & 32 & 86 \\
 & ANDROID & 305 & 3,582 & 2 & 8 \\
 & BATCH & 1,732 & 10,071 & 33 & 335 \\
 & BATCHADM & 243 & 1,298 & 4 & 16 \\
 & DATACMNS & 604 & 4,512 & 30 & 104 \\
 & DATAGRAPH & 848 & 5,190 & 14 & 43 \\
 & DATAJPA & 330 & 2,002 & 32 & 107 \\
 & DATAMONGO & 622 & 6,703 & 40 & 209 \\
 & DATAREDIS & 551 & 9,488 & 15 & 44 \\
 & DATAREST & 414 & 2,183 & 23 & 89 \\
 & LDAP & 566 & 3,556 & 5 & 46 \\
 & MOBILE & 64 & 814 & 3 & 8 \\
 & ROO & 1,109 & 7,803 & 15 & 568 \\
 & SEC & 1,618 & 9,295 & 41 & 422 \\
 & SECOAUTH & 726 & 3,912 & 6 & 61 \\
 & SGF & 695 & 5,790 & 19 & 83 \\
 & SHDP & 1,102 & 6,348 & 8 & 37 \\
 & SHL & 151 & 749 & 2 & 6 \\
 & SOCIAL & 212 & 1,344 & 4 & 10 \\
 & SOCIALFB & 253 & 1,786 & 4 & 11 \\
 & SOCIALLI & 180 & 830 & 1 & 2 \\
 & SOCIALTW & 153 & 1,197 & 5 & 6 \\
 & SPR & 6,512 & 57,696 & 10 & 89 \\
 & SWF & 808 & 6,864 & 19 & 101 \\
 & SWS & 925 & 3,505 & 24 & 122 \\ \hline
 & Total & 25,966 & 211,581 & 479 & 3,344 \\ \hline
  \end{tabular}}
\end{table}

\begin{table*}[tb]\centering
  \caption{Results of MAP}\label{t:map}
  \begin{tabular}{l|ccccc|ccccc} \hline
                   & \multicolumn{5}{c|}{File level}            & \multicolumn{5}{c}{Method level} \\
     Project & \BLT{BugLocator} & \BLT{BLUiR} & \BLT{BRTracer} & \BLT{AmaLgam} & \BLT{BLIA} & \BLT{BugLocator} & \BLT{BLUiR} & \BLT{BRTracer} & \BLT{AmaLgam} & \BLT{BLIA} \\\hline
        CODEC &\B0.631 & 0.623 & 0.283 & 0.629 & 0.621 & 0.192 & 0.352 & 0.094 & 0.345 & \B0.381 \\
        COLLECTIONS & 0.572 & 0.604 & 0.283 & 0.585 &\B 0.614 & 0.366 & 0.369 & 0.094 & 0.377 & \B0.394 \\
        COMPRESS & 0.631 &\B 0.703 & 0.263 & 0.678 & 0.696 & 0.281 & 0.318 & 0.170 & 0.302 & \B0.349 \\
        CONFIGURATION & 0.715 & 0.735 & 0.250 & 0.734 &\B 0.776 & 0.210 & 0.329 & 0.138 & 0.313 & \B0.363 \\
        CRYPTO & 0.314 & 0.313 & 0.063 & 0.322 & \B0.323 & 0.106 & 0.060 & \B0.264 & 0.060 & 0.058 \\
        CSV & 0.806 &\B 0.833 & 0.482 &\B 0.833 &\B 0.833 & 0.155 & 0.423 & 0.154 & 0.437 & \B0.453 \\
        IO & 0.742 & 0.761 & 0.271 & 0.764 & \B0.772 & 0.400 & \B0.506 & 0.227 & 0.499 & 0.495 \\
        LANG & 0.696 & 0.710 & 0.277 & 0.707 & \B0.738 & 0.374 & 0.503 & 0.167 & 0.525 & \B0.536 \\
        MATH & 0.499 & 0.553 & 0.145 & 0.571 &\B 0.578 & 0.263 & 0.355 & 0.122 & 0.355 & \B0.375 \\
        WEAVER & \B0.321 & 0.079 & 0.125 & 0.079 & 0.167 & 0.068 & \B0.083 & 0.015 & \B0.083 & 0.038 \\ \hline
        ENTESB & 0.119 &\B 0.431 & 0.399 & 0.429 & 0.329 & 0.031 & 0.167 &\B 0.350 & 0.166 & 0.214 \\
        JBMETA &\B 0.359 & 0.278 & 0.150 & 0.318 & 0.315 & 0.146 &\B 0.226 & 0.035 &\B 0.226 & 0.214 \\ \hline
        AMQP & 0.404 & 0.421 & 0.086 & 0.422 &\B 0.434 & 0.155 & 0.167 & 0.106 & 0.165 &\B 0.187 \\
        ANDROID & 0.540 & 0.562 &\B 0.566 & 0.499 & 0.556 & 0.298 & 0.343 & 0.237 & 0.353 &\B 0.367 \\
        BATCH & 0.414 & 0.436 & 0.122 &\B 0.448 &\B 0.448 & 0.216 & 0.222 & 0.138 & 0.227 &\B 0.241 \\
        BATCHADM & 0.438 & 0.553 & 0.276 & 0.549 &\B 0.606 &\B 0.256 & 0.223 & 0.136 & 0.248 & 0.225 \\
        DATACMNS & 0.300 & 0.365 & 0.136 &\B 0.369 & 0.362 & 0.125 & 0.204 & 0.124 & 0.201 &\B 0.214 \\
        DATAGRAPH & 0.145 & 0.171 & 0.107 & 0.182 &\B 0.197 & 0.145 & 0.171 & 0.106 & 0.182 &\B 0.197 \\
        DATAJPA & 0.355 & 0.369 & 0.110 & 0.380 &\B 0.394 & 0.169 & 0.173 & 0.115 & 0.175 & \B0.183 \\
        DATAMONGO & 0.296 & 0.317 & 0.111 & 0.316 &\B 0.340 & 0.114 & 0.144 & 0.096 & 0.142 &\B 0.165 \\
        DATAREDIS & 0.382 &\B 0.410 & 0.114 & 0.406 & 0.391 & 0.163 & 0.160 & 0.123 & 0.152 &\B 0.190 \\
        DATAREST & 0.264 & 0.272 & 0.119 &\B 0.290 & 0.289 & 0.100 & 0.127 & 0.077 &\B 0.128 &\B 0.128 \\
        LDAP & 0.498 & 0.476 & 0.223 & 0.487 &\B 0.508 & 0.227 & 0.300 & 0.212 &\B 0.309 &\B 0.309 \\
        MOBILE &\B 0.530 & 0.427 & 0.251 & 0.427 & 0.427 & 0.372 &\B 0.627 & 0.376 &\B 0.627 & 0.595 \\
        ROO &\B 0.414 & 0.375 & 0.127 & 0.384 &\B 0.414 & 0.200 & 0.193 & 0.087 & 0.200 &\B 0.228 \\
        SEC & 0.505 & 0.533 & 0.221 & 0.545 &\B 0.559 & 0.299 & 0.334 & 0.186 & 0.339 &\B 0.362 \\
        SECOAUTH & 0.434 & 0.426 & 0.157 & 0.429 &\B 0.457 & 0.301 & 0.275 & 0.176 & 0.285 &\B 0.318 \\
        SGF &\B 0.415 & 0.380 & 0.155 & 0.384 & 0.404 &\B 0.182 & 0.159 & 0.123 & 0.168 & 0.169 \\
        SHDP & \B0.387 & 0.361 & 0.187 & 0.360 & 0.379 & 0.200 & 0.161 & 0.166 & 0.160 &\B 0.211 \\
        SHL & 0.503 &\B 0.593 & 0.217 &\B 0.593 &\B 0.593 & 0.227 &\B 0.327 & 0.226 &\B 0.327 & 0.314 \\
        SOCIAL &\B 0.692 & 0.570 & 0.292 & 0.600 & 0.684 & 0.316 & 0.499 & 0.406 & 0.499 &\B 0.501 \\
        SOCIALFB &\B 0.666 & 0.470 & 0.222 & 0.495 & 0.520 & 0.184 & 0.300 &\B 0.328 & 0.318 & 0.318 \\
        SOCIALLI &\B 0.327 & 0.300 & 0.084 & 0.300 & 0.303 & 0.511 & 0.514 & 0.505 & 0.514 &\B 0.515 \\
        SOCIALTW & 0.693 & 0.462 &\B 0.694 & 0.462 & 0.512 &\B 0.488 & 0.320 & 0.193 & 0.320 & 0.282 \\
        SPR &\B 0.415 & 0.277 & 0.119 & 0.321 & 0.339 & 0.203 & 0.167 & 0.141 & 0.227 &\B 0.254 \\ 
        SWF & 0.461 & 0.424 & 0.253 & 0.411 &\B 0.476 & 0.212 & 0.243 & 0.191 & 0.229 &\B 0.271 \\
        SWS & 0.270 & 0.257 & 0.105 & 0.264 &\B 0.285 & 0.272 & 0.256 & 0.095 & 0.265 &\B 0.283 \\ \hline
   \end{tabular}
\end{table*}

Table \ref{t:cost} shows that the modifications of the two techniques were minimal (0.038 \% each).
Conversely, there were no modifications to techniques that use information directly related to files, such as file size and historical information.

\Conclusion{Our approach could convert five existing bug localization techniques to method level in eight lines.
The amount of modification was small, and it was easy to identify the required changes.}

\subsection{\RQ{2}: \RQtwo}

\Heading{Motivation}
We evaluated modified techniques in a large-scale unified framework to identify changes in performance to gain new insights into method-level bug localization.

\Heading{Study Design}
We compared the performance of the techniques at the same and different levels.
We evaluated the accuracy and effort required to determine the bug locations.
MAP and MRR were used as measures of accuracy, and the top-$k$ LOC\cite{Chakkrit2018IR} was used as a measure of effort.
Top-$k$ LOC is the percentage of bug reports for which at least one buggy file is found in the top-ranked files with a cumulative sum of LOC below $k$.
For example, a top-$k$ LOC of 0.1 for $k={}$10,000 means that the correct file appears within top 10,000 lines for 10\% of bug reports.
We used $k \in \{100,500,1000,5000\}$.
We used a two-sided Wilcoxon signed-rank test to evaluate the output of the techniques at both levels.

\Heading{Construction of FinerBench4BL Dataset}\label{ss:dataset}
In this experiment, we used 37 of the 46 projects provided by Bench4BL to construct the FinerBench4BL dataset owing to the execution time.
We filtered out bug reports that 1) any of the five IRBL techniques at either level failed to produce any solution.
A typical example case to be filtered out is that the code location to be fixed when resolving the bug was outside of methods, and the list of solution modules at the method level became empty.
The target projects, number of modules, versions, and bug reports to be used from Bench4BL and to be generated as FinerBench4BL are shown in Table~\ref{t:subject}.
25,966 original source files and 211,581 method files from all 479 versions were searched for 3,344 bug reports.

\def\WD{7.7cm}
\begin{figure}\centering{\footnotesize
  \includegraphics[width=\WD]{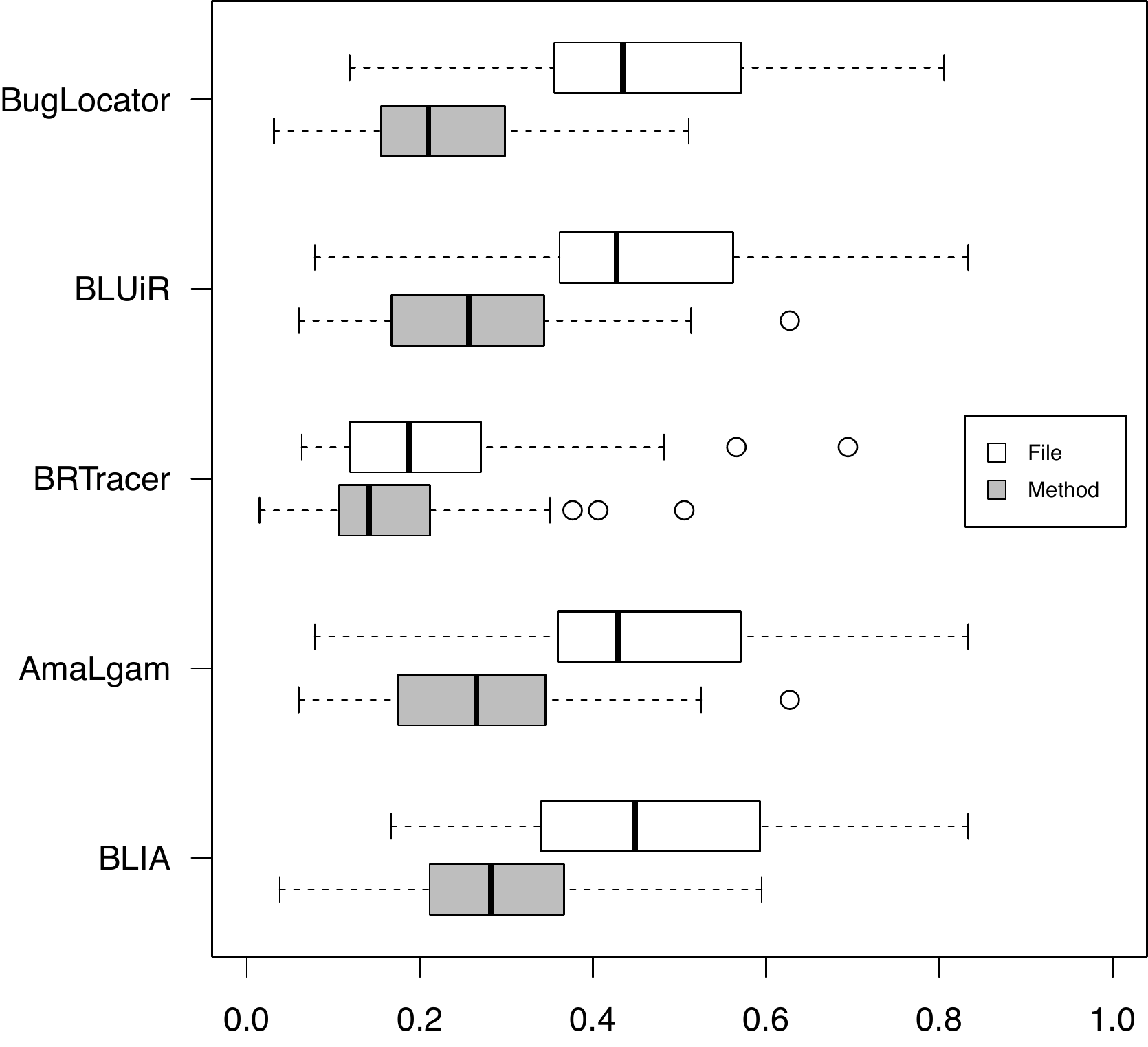}\\
  (a) MAP.\\~\\
  \includegraphics[width=\WD]{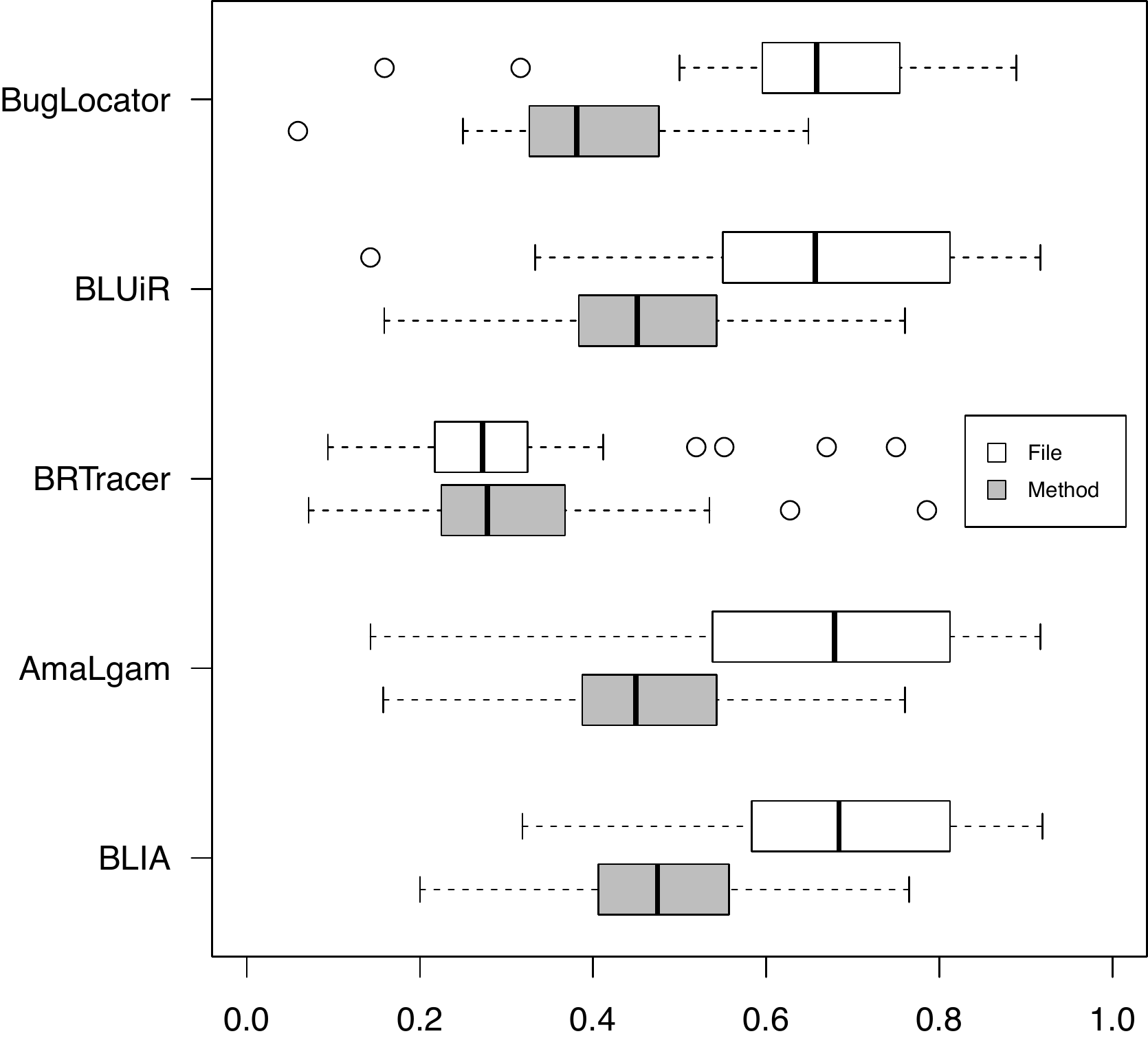}\\
  (b) MRR.\\~\\
  \includegraphics[width=\WD]{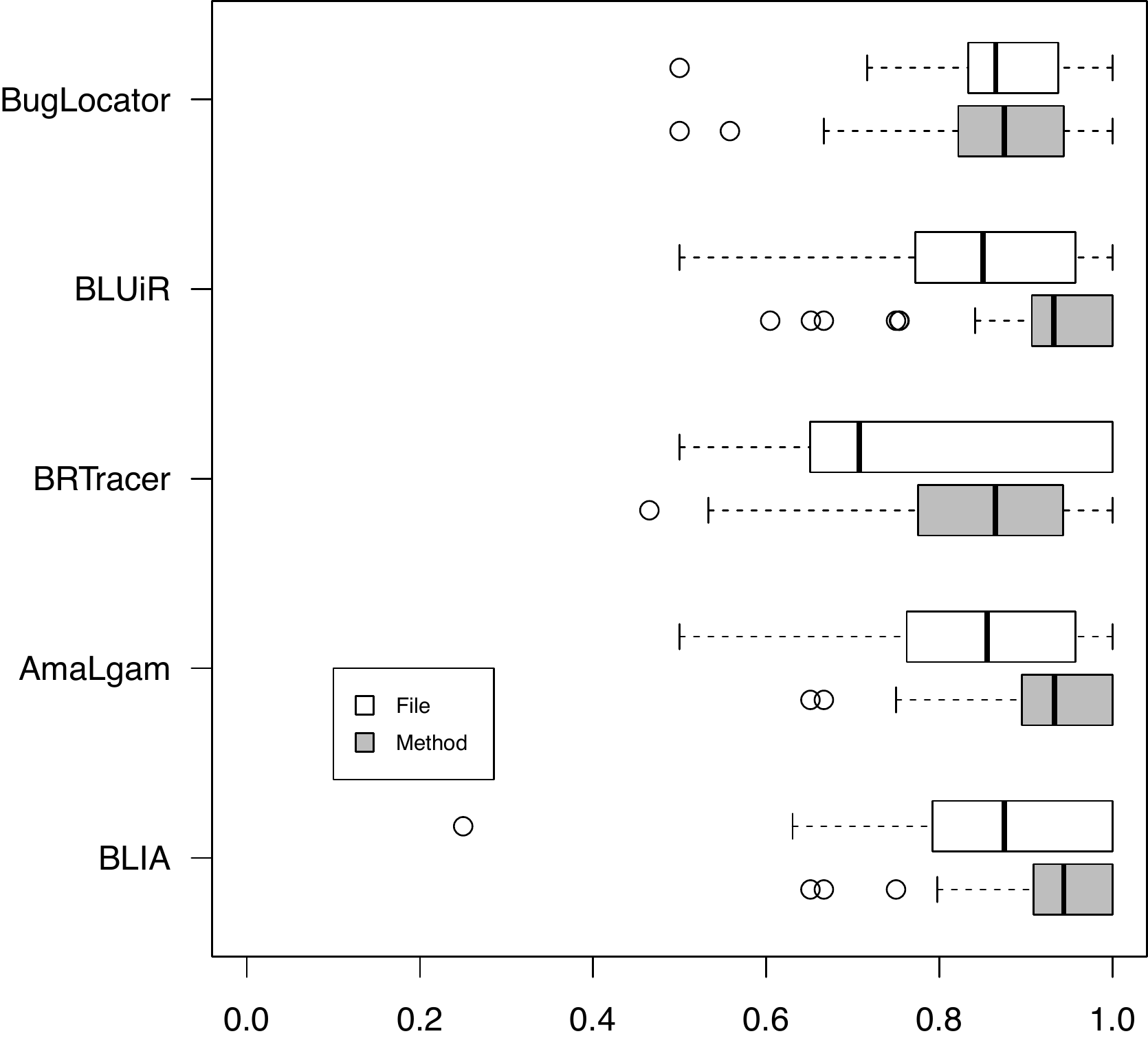}\\
  (c) Top-1,000 LOC.}
  \caption{Evaluation of IRBL techniques at each level.}\label{f:fm}
\end{figure}

\Heading{Results of Accuracy Evaluation}
Table~\ref{t:map} presents the MAP results for each project.
Box plots of the MAP and MRR of the projects comparing the file and method levels are shown in Figs.~\ref{f:fm}(a) and \ref{f:fm}(b).
We describe only the MAP results because the MRR results showed a similar tendency to the MAP results.
The technique with the highest accuracy for the same project is highlighted as bold in the results in Table~\ref{t:map}.

First, we investigated the accuracy at each level.
\BLT{BLIA} showed the highest accuracy, with the highest values in 20 of the 37 projects at the file level and 26 at the method level.

Furthermore, the same techniques showed the highest MAP at both levels for 18 projects (48.6\% of the total).
This means that, for each project, the technique that showed the highest accuracy at the file level is likely to recommend buggy methods precisely at the method level.
This result suggests that the selection of effective techniques for file-level bug localization may also be helpful at the method level.

Second, we investigated how the accuracy of each technique changed as it was modified for the method level.
There were significant differences in the median MAP between the techniques.
MAP at the file level was 1.33 times higher than that at the method level for \BLT{BRTracer} as the minimum and 2.07 times higher for \BLT{BugLocator} as the maximum.
Furthermore, at both levels, the median MAP value indicated by \BLT{BLIA} was the largest, followed by \BLT{AmaLgam}, \BLT{BLUiR}, \BLT{BugLocator}, and \BLT{BRTracer}.
Therefore, the accuracy increased with the amount of additional information handled, except for \BLT{BRTracer}.
The high accuracy of \BLT{BLIA}, which uses most of the information, indicates that it can maintain accuracy by supplementing the missing information at method-level bug localization from various perspectives.
In contrast, \BLT{BLUiR}, which uses only structural information as additional information, showed an accuracy comparable to that of \BLT{AmaLgam} and \BLT{BLIA}, which also consider other information.
This suggests that structural information is adequate for method-level bug localization.

In some bugs, significant changes in accuracy occur with different bug localization granularities.
Two examples of \BLT{BugLocator} outputs are presented below.

The first case is of decreased accuracy.
In \Bug{COLLECTIONS-220}\footnote{https://issues.apache.org/jira/browse/COLLECTIONS-220}, the \Mod{writeObject()} in \Mod{\seqsplit{UnboundedFifoBuffer.java}} was the cause of the bug.
Although the target file was ranked first at the file level, \Mod{writeObject()} was recommended in 789th place at the method level.
The words ``increment'', ``tail'', and ``head'' in this bug report were not included in any \Mod{writeObject()}, which was a deficient information method for the seven lines.
However, \Mod{add()}, \Mod{remove()}, and other methods included in \Mod{UnboundedFifoBuffer.java}, contained several relevant words.
This suggests that information from irrelevant methods contributes to the high accuracy in file-level bug localization.

The second case is of improved accuracy.
In \Bug{CONFIGURATION-558}\footnote{https://issues.apache.org/jira/browse/CONFIGURATION-558}, \Mod{getList()} in \Mod{\seqsplit{MultiFileHierarchicalConfiguration.java}} was the buggy method.
The method-level bug localization improved the rank from 22nd to seventh at the file level.
Although there were few bug report descriptions, they included the parameters and method names directly related to the target method.
These entities might have contributed to the improvement in the accuracy at the method level.
This is supported by the results of Wang \etal~\cite{Wang2015BR} and Rahman \etal~\cite{Mohammad2018QueryReform}, who showed that bug reports containing program entities are suitable for IRBL.

These examples with significant variations in accuracy at both levels suggest that, at the file level, information on methods irrelevant to the bug location is used for information retrieval.
This suggests the output ranked list based on unrelated methods may not help developers find such bug locations.
Conversely, the textual information of the method body may be insufficient for information retrieval at the method level, which is considered to have reasonable granularity.
Therefore, for method-level IRBL, it is useful to implement a hybrid approach for bug localization that uses not only the information on the target method body but also the information outside of its own method, as proposed by Amasaki \etal~\cite{Amasaki2020}.

\Conclusion{\BLT{BLIA} showed the highest accuracy at both levels.
In 48.6\% of the projects, the best-performed techniques were the same at both levels.
Therefore, an accurate technique at the file level performs well at the method level.
Furthermore, we found that the additional information effectively contributed to the recommendation of the buggy methods despite a 2.07-fold difference in accuracy.}

\Heading{Results of Effort Evaluation}
Figure~\ref{f:fm}(c) shows the results of the top-1,000 LOC, which is a measure of effort to find bug locations from the ranked list.
Table~\ref{t:topkloc_median} also presents the top-1,000 LOC median, p-value of the two-sided Wilcoxon signed-rank test, and Cliff's $d$ with an interpretation~\cite{romano-fair2006} for each project.
Owing to space limitations, we omit the results using $k \in \{100,500,5000\}$ because they produce similar tendencies to the case of $k=1000$.

\begin{table}[tb]\centering
  \caption{Median of Top-1,000 LOC Values}\label{t:topkloc_median}
  {\tabcolsep=5.5pt\begin{tabular}{l|cc|cl} \hline
   & File level & Method level & p-value & \multicolumn{1}{c}{Cliff's $d$} \\ \hline
    \BLT{BugLocator} & 0.865 & 0.875 & 0.644 & 0.063 (negligible) \\
    \BLT{BLUiR} & 0.850 & 0.933 & 0.026 & 0.297 (small) \\
    \BLT{BRTracer} & 0.708 & 0.865 & 0.024 & 0.301 (small) \\
    \BLT{AmaLgam} & 0.855 & 0.933 & 0.022 & 0.305 (small) \\
    \BLT{BLIA} & 0.875 & 0.944 & 0.027 & 0.295 (small) \\ \hline
   \end{tabular}}
\end{table}

For all techniques, top-1,000 LOC performance was improved at the method level.
Among the four techniques, except for \BLT{BugLocator}, there were significant differences in the top-1,000 LOC between the two levels.

Compared to the top-1,000 LOC performance at the file level, \BLT{BugLocator} showed the smallest increase at 2.2\%, and \BLT{BRTracer} showed the largest increase at 22.2\%.
At the method level, the best performing \BLT{BLIA} identified correct files within the top-1,000 lines of code of the ranked list for 94.4 \% of the bug reports.

The top-1,000 LOC performance of the file level was lower than that of the method level owing to the large size of each file's lines of code, despite better accuracy.
This suggests that the number of lines developers need to read could be reduced by current method-level IRBL approaches, even if their accuracy is not high.

In terms of the performance order, \BLT{BLIA} showed the highest top-1,000 LOC, and \BLT{BRTracer} showed the lowest, at both levels.
The order of the method level performance of the top-1,000 LOC for each technique is the same from the file level, as with the results of the accuracy evaluation, except for \BLT{BugLocator}, which did not show significant differences.
This indicates that a technique with a good effort performance at the file level also performed well at the method level.
Furthermore, as in the above discussion of accuracy, \BLT{BLUiR}, which only uses structural information, performed top-1,000 LOC as well as the other techniques at both levels.
Therefore, the consideration of structural information leads to an improvement in the top-1,000 LOC performance.

\Conclusion{Except for one technique, the method-level techniques significantly improved the top-1,000 LOC\@.
\BLT{BLIA}, which best performed, found fixed method files in more than 94\% of bug reports within the top 1,000 lines of code of the rank list.
Techniques that perform well at the file level often also perform well at the method level.
Structural information may improve the performance in method-level bug localization, where information is scarce.}

%%%%%%%%%%%%%%%%%%%%%%%%%%%%%
\section{Threats to Validity}\label{s:validity}

In this study, we assumed that bug reports were correctly linked with fixed files.
We used scripts provided by Bench4BL, linked bug reports, and fixed files.
However, we did not check the validation of the correctness of the linking; therefore, there may be bug reports that cannot be identified in the commit log and those linked with non-buggy files.

Although the top-$k$ LOC was adopted as an indicator for evaluating the effort, its validity to the actual effort required to identify bug locations was not apparent.
In addition, we used the total number of LOC from the top of the ranked list above the correct solution file as an indicator of effort.
However, if the solution is ranked at the top, then the total number is zero.
This result may differ from the actual effort.
Moreover, this indicator also affects the evaluation in terms of effort as developers do not necessarily read all lines of files.

%%%%%%%%%%%%%%%%%%%%%%%%%%%%%

\section{Conclusion}\label{s:conclusion}

In this study, we propose an approach to convert IRBL techniques to the method level using repository transformation.
We evaluated them at both method and repository levels using a framework combining converted techniques with datasets constructed at the method level.
This approach can convert the existing IRBL techniques to method-level techniques with minor modifications.

The evaluation results showed that the converted method-level techniques decreased the accuracy but reduced the debugging effort.
However, as Amasaki \etal~\cite{Amasaki2020} stated, there is potential for further performance improvement using file-level information to compensate for the lack of details in method-level bug localization.
In future research on method-level IRBL techniques, obtaining performance improvements will be easier using our framework to adjust the parameters and select additional information.

The future research directions of this study are as follows:
\begin{itemize}
  \item \emph{Increasing the number of evaluated techniques and datasets.}
  We believe the evaluation framework can be extended to existing method-level techniques and other datasets to gain more knowledge.
  \item \emph{Tuning the parameters of method-level techniques.}
  The method-level techniques in this paper are run with parameters tuned for the file level.
  Further performance improvements could be achieved by identifying optimal settings for method-level bug localization.
  \item \emph{Comparing both levels of techniques under more uniform conditions.}
  When it is necessary to find all methods in a fixed file, the number of solutions that method-level techniques need to recommend is higher than at the file level, contributing to a significant reduction in accuracy.
  Bug reports with many solution methods should be excluded and compared under appropriate conditions.
  \item \emph{Considering code elements outside of methods.}
  Incorporating information in Java source files, such as fields and imports, may provide different insights, which were excluded in this study.

\end{itemize}
%%%%%%%%%%%%%%%%%%%%%%%%%%%%%

The FinerBench4BL experimental results of this study are publicly available~\cite{zenodo}.

\section*{Acknowledgments}
This study was partially supported by JSPS KAKENHI (JP21H04877, JP21K18302, JP21KK0179, 21K11833, and JP22H03567).

\IEEEtriggeratref{10}

% Generated by IEEEtran.bst, version: 1.12 (2007/01/11)

\end{document}